\documentclass[usenatbib]{mn2e}

\usepackage{aas_macros}
\usepackage{astro-paper}
\usepackage{xcolor}
\usepackage{times}

\begin{document}
\title[Jet Ejection in Common Envelope Interactions]{Constraints on Common Envelope Magnetic Fields from Observations of Jets in Planetary Nebulae}
\author[Tocknell et al.] {James Tocknell, Orsola De Marco, Mark Wardle\\ Department of Physics and Astronomy, Macquarie University, Sydney, NSW 2109, Australia\\ Astronomy, Astrophysics and Astrophotonics Research Centre, Macquarie University, Sydney, NSW 2109, Australia}

\maketitle
\begin{abstract}
The common envelope (CE) interaction describes the swallowing of a nearby companion by a growing, evolving star. CEs that take place during the asymptotic giant branch phase of the primary may lead to the formation of a planetary nebula (PN) with a post-CE close binary in the middle. We have used published observations of masses and kinematics of jets in four post-CE PN to infer physical characteristics of the CE interaction. In three of the four systems studied, Abell~63, ETHOS~1 and the Necklace PN, the kinematics indicate that the jets were launched a few thousand years before the CE and we favour a scenario where this happened before Roche lobe overflow, although better models of wind accretion and wind Roche lobe overflow are needed. The magnetic fields inferred to launch pre-CE jets are of the order of a few Gauss. In the fourth case, NGC~6778, the kinematics indicate that the jets were launched about 3000 years {\it after} the CE interaction.  Magnetic fields of the order of a few hundreds to a few thousands Gauss are inferred in this case, approximately in line with predictions of post-CE magnetic fields. However, we remark that in the case of this system, we have not been able to find a reasonable scenario for the formation of the two jet pairs observed: the small orbital separation may preclude the formation of even one accretion disk able to supply the necessary accretion rate to cause the observed jets. 
\end{abstract}

\begin{keywords}
  Magnetic Fields, ISM: Jets and Outflows, Planetary Nebulae: Individual: Necklace, Planetary Nebulae: Individual: Abell 63, Planetary Nebulae: Individual: ETHOS 1, Planetary Nebulae: Individual: NGC 6778
\end{keywords}

\section{Introduction}
A common envelope (CE) interaction between a giant and a more compact companion happens when the envelope of the primary giant star grows sufficiently large as to engulf the secondary. The orbital energy unbinds the envelope, leaving either a close binary composed of a white dwarf (the core of the giant) and the companion or, if the secondary lacks sufficient energy to unbind the envelope, a merger \citep{Paczynski1976, ivanova_common_2013}. 

The shapes of most planetary nebulae (PN) diverge significantly from spherical \citep{parker_macquarie/aao/strasbourg_2006}. The reason may be that a companion has played a role during the mass-losing asymptotic giant branch (AGB) phase during which the PN gas was ejected \citep{Soker1997,DeMarco2009}. While for most PN with non-spherical shapes we presume the action of a companion, for approximately one in five PN {\it we know} that a companion has ejected the envelope \citep{bond_binarity_2000,miszalski_binary_2009}. Common envelope PN are identified by the presence of a close binary in the centre of a PN. On occasion a PN surrounding a post-CE central binary turns out to be a Stroemgren sphere around a post-red giant branch star, rather than a proper PN (\citealt{Frew2010c}; e.g., EGB~5; \citealt{Geier2011}). PN around post-CE binaries are not only interesting because they are cases for which we know the mechanisms that imparted the PN it's shape \citep[e.g.][]{Miszalski2009b}, but also because they provide a unique tool for the study of the CE interaction. In post-CE PN the existence and brightness of the PN guarantees that the CE interaction only took place a few thousand years ago at most. Also, the aftermath of the ejection is there to be studied.

In this paper we focus on four PN around post-CE binaries, which exhibit jet-like structures. These have been measured and their kinematics indicate that  three of the four objects launched their jets before the main nebula, a clear indication that an accretion disk formed before the companion plunged into the primary. In the fourth case not one, but two pairs of jets are observed. Both these jet pairs are kinematically younger than the nebula, demonstrating that they were launched after the CE interaction had taken place. 

We assume here that all CE jets are launched via an accretion disk threaded by a magnetic field \citep{Blandford1982}. The launch efficiency, or the fraction of accreted mass that is ejected, is between 10\% and 50\% in this model \citep{sheikhnezami_bipolar_2012}. We also assume that the magnetic field responsible for the jet launching and collimation is also responsible for the loss of angular momentum and accretion of disk material \citep{wardle_magnetic_2007}. Within this model, we use the jet properties to determine the magnetic field responsible for the mass accretion and jet launching in pre and post-CE phases. 

In \secref{data} we review the four PN which will be analysed in this paper, outlining the kinematics and morphology of these objects. \Secref{AccretionAndEjectionMechanisms} contains our examination of a number of different scenarios for launching the jets. In \Secref{magfields} we discuss the magnetic fields that are implied for the four PN under the assumptions of our models and compare them, in the case of the post-CE PN, to the magnetic fields theoretically derived in a model by \citet{regos_effect_1995}. In \Secref{NGC6778Model}, we discuss scenarios for the formation of the post-CE jets of NGC~6778 and in \Secref{Comparison} we briefly discuss two additional post-CE PN. Finally, in \Secref{Conclusion}, we summarise and conclude.

\section{Post-CE systems with jets}\seclabel{data}
A few PNe with jets are known to contain post-CE central star binaries, but in only four cases have we sufficient information to carry out our study: the Necklace \citep[PN G054.2--03.4;][]{corradi_necklace:_2011}, Abell 63 \citep[
PN G053.8--03.0, whose central stars is known as UU~Sagittae;][]{mitchell_proof_2007,afsar_two-colour_2008}, ETHOS~1 \citep[PN G068.1+11.0;][]{miszalski_ethos_2011} and NGC~6778 \citep[PN G034.5--06.7;][]{guerrero_ngc_2012,miszalski_discovery_2011}. In order to develop a model for launching the observed jets, we need values for the kinematic parameters and masses of the circumstellar material. These are summarised in \Tabref{data}. 

\begin{table*}
  \begin{threeparttable}[t]
  \caption{Properties of the common envelope PN in our sample}
  \tablabel{data}
  \begin{tabular}{lllll}
    Object                                                      & \raggedright The Necklace             & Abell 63                & ETHOS 1             & NGC 6778                    \\
    \hline
    Deprojected expansion velocity of main nebula/ring ($\kms$) & $28 \pm 3$                            & $\sim 26$               & $\sim 55$           & $\sim 26$                   \\
    Deprojected velocity of jets ($\kms$)                       & 95(N), 115(S)\tnote{a}                & $126 \pm 23$\tnote{a}   & $120\pm10$\tnote{d} & 270\tnote{e}, 460\tnote{f}  \\
    Ionised gas mass of nebula/ring ($\Msun$)                   & $0.06\pm0.03$                         & $0.09$                  & --                  & --                          \\
    Ionised gas mass of jets ($\Msun$)                          & $\sim 10^{-3}$                        & --                      & --                  & $\sim 1.5 \times 10^{-3}$\tnote{g}     \\
    Radius of nebula/ring ($\uarcsec$)                          & $6.5 \pm 0.5$                         & $\sim 14$               & $\sim 9.7$          & $\sim 8.5$                  \\
    Distance to jet tips from star ($\uarcsec$)                 & $\sim 60$                             & $\sim 142$              & $\sim 31.3$         & $\sim 35$                   \\
    Age of nebula/ring ($\uyear$)                               & $\sim 5000$                           & $\sim 11200$            & $\sim 5400$         & $\sim 4400$                 \\
    Age of jets ($\uyear$)                                      & $\sim 8700\tnote{b} - 13000\tnote{c}$ & $\sim17100$             & $\sim 10500$        & $\sim 1700$                 \\
    Jet ejection timescale (yr)                                 & $3700-8000$                           & 5900                    & 5100                & $1700$                     \\
    Jet mass-loss rate ($\Msun\usk\reciprocal\uyear$)           & $1 - 3\times 10^{-7}$                 & --                      & --                  & $8.8 \times 10^{-7}$    \\
    Period of central binary ($\text{days}$)                    & 1.2                                   & 0.46                    & 0.53                & 0.15                        \\
    Distance to object ($\kpc$)                                 & $4.6\pm1.1$                           & $3.2\pm0.6$             & $6.0^{+2.5}_{-1.5}$ & $2.6^{+0.7}_{-0.8}$         \\
    \hline
  \end{tabular}
  \begin{tablenotes}
    \item [a] Average velocity of caps.
    \item [b] Edge of southern cap closest to central star.
    \item [c] Edge of southern cap furthest from central star.
    \item [d] Velocity of jet tips.
    \item [e] Linear jet.
    \item [f] Curved jet.
    \item [g] Mass of each pair.
  \end{tablenotes}
  \end{threeparttable}
\end{table*}

\subsection{The Necklace}
\seclabel{TheNecklace}

The Necklace PN (PN G054.2--03.4; Fig.~1, left panel) consists of a ring with radius $\unit{6.5\pm0.5}{\uarcsec}$ expanding at $\unit{28 \pm 3}{\kms}$ on a plane inclined by $\unit{59 \pm 3}{\degree}$ to the line of sight, where $\unit{0}{\degree}$ is in the plane of the sky \citep{corradi_necklace:_2011}. Two polar caps are assumed to be perpendicular to the plane of the ring. The northern cap is quite compact and spans a distance of 38 to $\unit{46}{\uarcsec}$ from the star. The southern cap is more extended and spans a distance of 37 to $\unit{59}{\uarcsec}$ (measured by us on figure 4 of \citet{corradi_necklace:_2011}). All the gas in each cap appears to move with the same velocity:  deprojected velocities are $\unit{95\mathand115}{\kms}$, for the northern and southern caps, respectively. We gauged the errors on these velocities to be $\unit{3}{\kms}$ from figure 4 of \citet{corradi_necklace:_2011}. The kinematic ages are determined by dividing  the ring radius (or the deprojected distance between the star and the base or tip of each jet cap) by the deprojected ring expansion velocity (or deprojected velocity of the caps). For the ring, \citet{corradi_necklace:_2011} find a kinematic age of $\unit{1100}{\yrkpc}$. The error on this estimate can be estimated to be 25\%. The kinematic age of the innermost part of the southern cap is $\unit{1900}{\yrkpc}$, while for the outermost part it is $\unit{2800}{\yrkpc}$. The northern cap has an average age of $\unit{2500}{\yrkpc}$. The error on these estimates can be determined to be 20\%.  The lack of a velocity gradient along the caps implies that material was ejected during a certain period of time, as opposed to during a quick outburst. The distance to this PN was kindly measured by D. Frew to be $\unit{4.6\pm1.1}{\kpc}$ by applying the surface brightness-radius relation \citep{frew_planetary_2008}. At that distance the ring has an age of $\sim$5000 years while the  age of the polar caps would be in the range $8700-13\,000$ years. Using this distance estimate, the time over which the jet was launched is $\unit{\sim4000-8000}{\uyear}$, which agrees with the assertion that the ejection was not a quick outburst. 

\begin{figure*}
\vspace{8cm}
\includegraphics{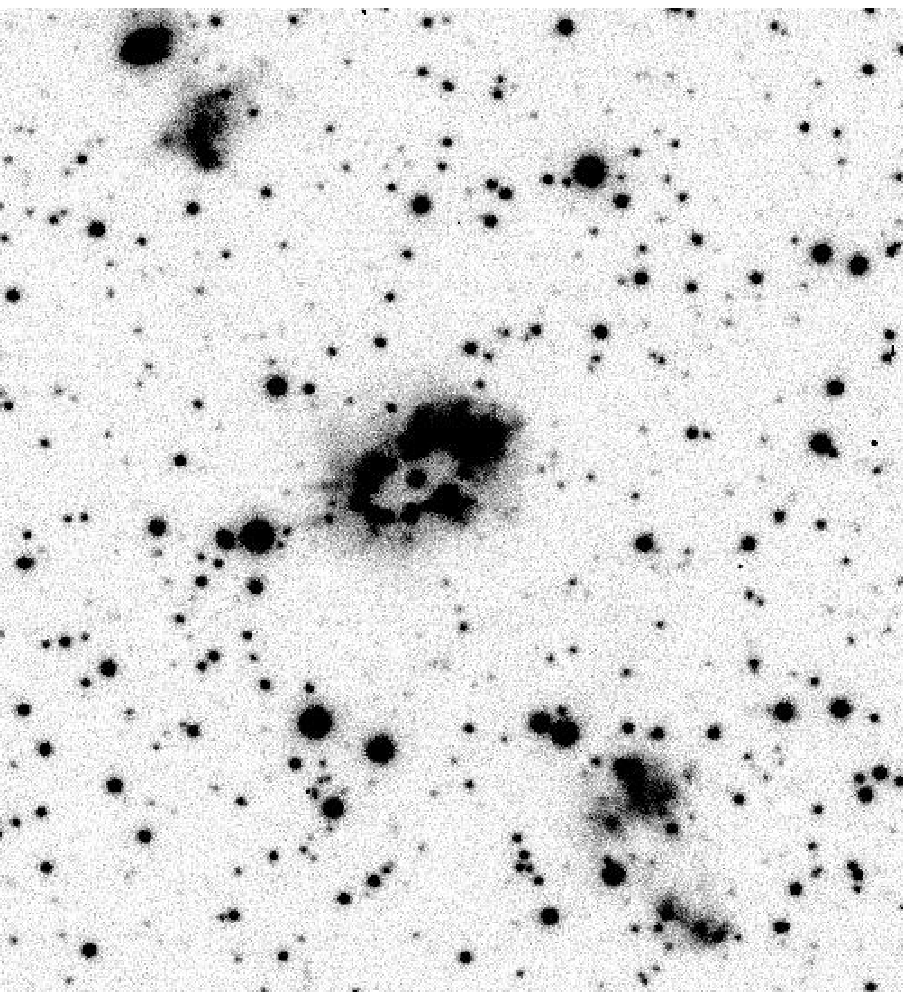}
\includegraphics{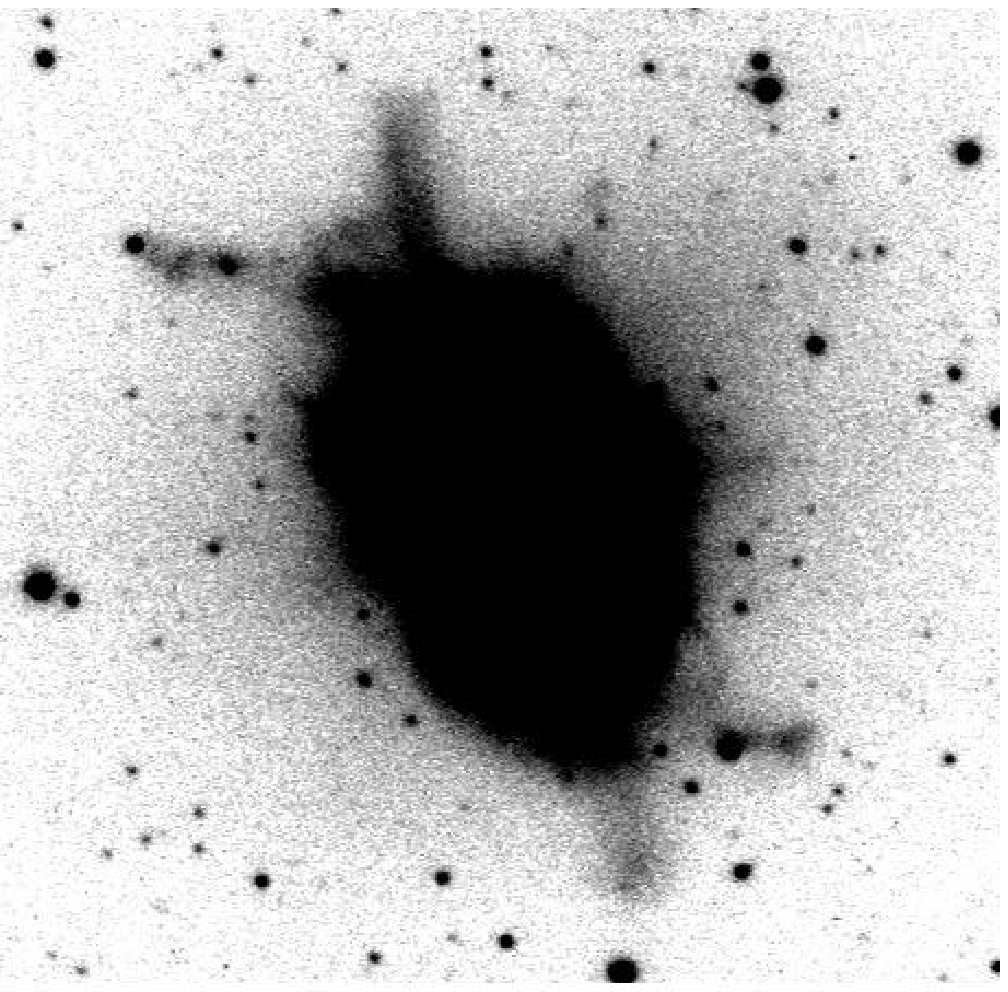}
\caption{Left panel: An [NII] image of the Necklace PN created using data from Corradi et al. (2011). The height of the image is approximately $\unit{100}{\uarcsec}$. Right panel: an $H_\beta$ image of NGC~6778 adapted using data from \citet{guerrero_ngc_2012}. The image's height is approximately 70 arc seconds. In both images, North is to the top and East is to the left. Both images were scaled so as to emphasise the jet structures. For additional PN structure details refer to the original papers}
\end{figure*}

The ionised mass of the entire nebula was estimated from the integrated $H_{\alpha}$ flux and a filling factor of 0.4 to be $\unit{0.06\pm0.03}{\Msun}$. The ionised mass of the caps is of the order of $\unit{10^{-3}}{\Msun}$ (R. Corradi, private communication). These mass estimates as well as the estimates of the kinematic ages determined above, will suffer from an additional source of uncertainty, which is hard to quantify, but which we argue to be smaller than an order of magnitude. All ejected material impacts the circumstellar ambient medium. This will potentially increase the mass of the ejected structures and decrease their speeds over time. 

We argue here that it is unlikely that the PN body and jets have slowed down dramatically. The velocities measured along the cap (particularly the southern cap which spans a range of about $\unit{20}{\uarcsec}$) are surprisingly constant with no real indication that the material launched most recently is significantly faster (if anything the gas closest to the star is slowest). We also present circumstantial evidence that the measured velocities are not much smaller than the ejection velocities, and in any case not by more than a factor of a two: the jet velocities measured for the PN A63  (\Secref{Abell63}), ETHOS~1 (\Secref{ETHOS1}) and FLEMING~1 (\Secref{Comparison}) are all approximately $\unit{100}{\kms}$, something that would be unlikely were they slowed down considerably by material in four different circumstellar environments.

The momentum conservation considerations presented by \citet{blackman_magnetic_2009} could be applied in the current case to estimate the amount of mass loading and the velocity decrease. However, contrary to the case described by \citet{blackman_magnetic_2009}, where the {\it pre}-PN jets are punching through the entire AGB star envelope (which has just been ejected in the superwind phase), the present case is different. The common envelope is likely to take place before the super wind ensues. We argue this on probabilistic grounds. The chance that the capture of a companion coincides with the very short, final, phase of the AGB star life is unlikely. It is more likely that at some point in the upper AGB, but before the superwind phase, the companion was captured. In such case the circumstellar material encountered by the jet would be far less dense than in the case of a jet perforating the super wind-formed shell, which is effectively the entire AGB envelope. Even wanting to follow through with the calculation of \citet{blackman_magnetic_2009}, the uncertainty in determining the swept up mass would outweigh that of assuming that the jet mass today is the same as the ejected mass, which, we argued above, should only be within a factor of less than two.

Finally, we note that for the central star of the Neckalce PN, there is independent observational evidence that accretion has occurred onto the companion \citep{miszalski_carbon_2013}, because of the pronounced carbon abundance of this otherwise normal main sequence star. The estimated amount of accreted matter was $\unit{0.03-0.35}{\Msun}$ for a $\unit{1.0-0.4}{\Msun}$ companion. Inspection of the equations used by \citet{miszalski_carbon_2013}, reveals that in the case of a $\unit{0.3}{\Msun}$ main sequence companion, which is almost fully convective, we expect between 0.10 and $\unit{0.42}{\Msun}$ of gas accreted to raise the C/O ratio to unity from the solar value, for a range of AGB C/O ratios of $1.5-3$. In \Secref{RocheLobeOverflow} we will consider whether such a large accreted mass is in line with the deduced accretion rates and jet mass-loss rates.

\subsection{Abell 63}\seclabel{Abell63}
Abell 63 (PN G053.8--03.0) is a faint planetary nebula discovered by \citet{abell_properties_1966}, with the binarity of central star discovered by \citet{bond_uu_1978}. Like the Necklace, it appears as an edge-on ring with two caps. \citet{mitchell_proof_2007} provided detailed kinematic and morphological measurements of the PN. The inclination of the system was determined to be $\unit{87.5}{\degree}$ (where $\unit{0}{\degree}$ is in the plane of the sky), assuming that the inclination of the system is the same as the inclination of the binary.

\citet{mitchell_proof_2007} measured the expansion velocity of the ring or torus structure to be $\unit{\sim 26}{\kms}$. The average radial velocities of the caps are $\unit{5.5 \pm 1}{\kms}$, and when the inclination of the system is taken into account, the average velocities of the caps are $\unit{126 \pm 23}{\kms}$. \citet{frew_planetary_2008} measured the ionised mass of Abell 63 to be $\unit{0.09}{\Msun}$, using a filling factor of $0.4$, similar to the total PN mass of the Necklace nebula. No jet mass estimate exists for this object. \citet{mitchell_proof_2007} derived kinematic ages of the structures using a distance of $\unit{2.4}{\kpc}$, which they attributed to \citet{Pollacco1993}. However, those authors actually derived a distance of $\unit{3.2\pm0.6}{\kpc}$ from the eclipsing binary system, a directly-determined distance which we use in the present study. The surface brightness-radius relation of \citet{frew_planetary_2008} results in a smaller distance estimate of  $\unit{2.6^{+1.0}_{-0.5}}{\kpc}$ which is consistent with the eclipsing binary distance estimate within the large error bars. We have rescaled the dynamical ages of the nebular structures of Abell~63 to the distance of $\unit{3.2}{\kpc}$: $\sim 11\,200$~yr for the ring and $\sim 17\,100$~yr for the jets. These larger age values are in line with the appearance of Abell~63, characteristic of an old nebula. According to these figures, the jet predates the nebula by $\sim 5900$~yr. We gauge the errors on these estimates to be similar to those estimated for the Necklace PN case (\Secref{TheNecklace}). Finally, \Citet{mitchell_proof_2007} suggests that the ejection was fairly rapid, contrary to the case of the Necklace PN, because the morphology of the jets of Abell 63 match those of Mz 3, which is known to have had launched the jets over a short amount of time.

\subsection{ETHOS 1}
\seclabel{ETHOS1}

ETHOS~1 (PN G068.1+11.0) was discovered by \citet{miszalski_ethos_2011}, as part of a survey of the SuperCOSMOS Science Archive \citep{hambly_supercosmos_2004}. This nebula appears as a torus, with two perpendicular outflows. The angle of inclination of the disk was given as $\unit{60\pm5}{\degree}$ to the line of sight, where $0\degree$ is in the plane of the sky. The torus has radius of $\sim$10~arcsec and a radial velocity gradient with a maximum velocity of $\unit{55}{\kms}$, possibly implying a short timescale for the ejection. The SE jet has radial velocity $\unit{-55 \pm 5}{\kms}$ and the NW jet has radial velocity $\unit{65 \pm 5}{\kms}$. The deprojected velocities of the jets are $\unit{120 \pm 10}{\kms}$ \citep{miszalski_ethos_2011}, assuming that the jets are symmetric. \citet{miszalski_ethos_2011} give kinematic age calculations for both the jets, $\unit{1750 \pm 250}{\yrkpc}$, and for the inner nebula $\unit{900 \pm 100}{\yrkpc}$. There are no  distance estimates for ETHOS~1 in the literature. Using the surface brightness-radius relation, a distance of $6.0^{+2.5}_{-1.5}$~kpc can be derived (Frew 2008). With this distance the age of the jets is 10\,500~yr, and the age of the ring is 5400~yr, implying that the jets predate the nebula by $\sim5100$~yr. We estimate the errors on these estimates to be similar to those estimated for the Necklace PN case (\Secref{TheNecklace}). No mass estimates exist for this PN.

\subsection{NGC~6778}
\seclabel{NGC6778}

The PN NGC~6778 (PN G034.5--06.7; Fig.~1, right panel) was discovered to harbour a post-CE central star by \citet{miszalski_discovery_2011}. \citet{maestro_imaging_2004} and \citet{guerrero_ngc_2012} carried out detailed kinematical analysis of the nebula. The equatorial ring has an radius of $\unit{8.5}{\uarcsec}$. The ring has an inclination to the line of sight of $\unit{\sim 75-78}{\degree}$ (where $\unit{0}{\degree}$ is in the plane of the sky) and was observed to expand with a deprojected velocity of $\unit{26}{\kms}$. From the ring protrude two lobes extending approximately $\unit{20}{\uarcsec}$ from the centre of the nebula. The deprojected expansion velocity at the tip of the lobes is $\unit{50}{\kms}$. The authors note that the lobes lack a typical velocity structure, and suffer instead from great complexity, as if they had been bored along selected directions. 

The agent responsible for the shaping seems to be two pairs of collimated features extending farther than the lobes, to $\unit{\sim35}{\uarcsec}$ from the centre of the nebula. One pair is linear and is approximately aligned with the bipolar lobes. The second pair starts near the star with the same inclination but curves at the tips with point symmetry. Both pairs of jets exhibit a velocity gradient, with velocity increasing as the distance from the centre. Assuming the normal to the disk plane makes an angle of $\unit{78}{\degree}$ with the line of sight, the jets have deprojected velocities of $\unit{270}{\kms}$, for the linear jets and $\unit{460}{\kms}$, for the curved jets. The kinematic ages of the ring and lobes are $\unit{1700}{\yrkpc}$ and $\unit{1600}{\yrkpc}$, respectively, while for the linear jets it is $\unit{650}{\yrkpc}$. \textit{This clearly indicates that the jets of NGC~6778 were launched after the main nebula, contrary to those of the other three post-CE PN analysed here}. This is one of two post-CE PN for which the jets are kinematically younger than the main nebula (the other, NGC~6337, is described in \Secref{Comparison}).

The lack of any change in the images taken three years apart imposes a lower limit on the distance of $\unit{1}{\kpc}$. The 9 distance estimates listed in the ESO PN catalogue \citep{acker_properties_1992} range between $\unit{1.9 \mathand 3.1}{\kpc}$, with only one estimate at $\unit{8.1}{\kpc}$. The surface brightness-radius distance to this object is $2.6^{+0.7}_{-0.8}$~kpc (Frew 2008). Using this distance estimate we calculate that the jets are 1700 years old, while the main nebula is 4400 years old. In this nebula the jets lag the nebular ejection  by 2700~yr.

The mass of the jets was kindly obtained by M. Guerrero. The average $H_{\beta}$ surface brightness in the jets is $\unit{\sim 1.0 \times 10^{-16}}{\erg\usk\centi\metre\rpsquared\reciprocal\second\rpsquare\uarcsec}$, with an uncertainty of 15\% to account for the spectroscopic calibration and extinction correction uncertainties, and for the slit location on the nebula.  Then, assuming the jets to be ``cylinders'' of radius $\unit{2}{\uarcsec}$ and height $\unit{27}{\uarcsec}$ for the linear jets and $\unit{23}{\uarcsec}$ for the curved jets a root mean square density $N_{e} \sim \unit{110 \times \epsilon^{-0.5}}{\centi\metre\rpcubed}$ is obtained. This leads to masses of $\unit{1.2\times 10^{-4}\epsilon^{0.5} D^{2.5}}{\Msun}$ for each linear jet, and $\unit{9.9\times 10^{-5} \epsilon^{0.5}D^{2.5}}{\Msun}$ for each curved jet, where $D$ is the distance to the object in kiloparsecs and $\epsilon$ is the filling factor. If we adopt a filling factor of 0.4 and a distance of 2.6~kpc, we obtain jet masses of $\unit{1.6\times 10^{-3}}{\Msun}$ and $\unit{1.4\times 10^{-3}}{\Msun}$ for the linear and curved jet {\it pairs}, respectively.

The formal error in the determination of the ring radius and location of the caps from the spatio-kinematic analysis is approximately 2~arcsec, while the velocity error is due to the width of the line and can be (generously) determined to be 10~km~s$^{-1}$. The error in the inclination is approximately 3 deg. So the formal error on the distance-independent ages is approximately 40\%. Applying this error in opposite directions to the kinematic age of the jets and the ring, so as to reduce their difference, brings the two values to be within 100 yrs of one another. Although this would effectively indicate coevality of the structure, it is unlikely that the sequence of the ejection would be completely reversed (jets before nebula as is the case for the three PN described above). As we will point out later on, these jets have many differences to the ones just described but do have commonalities with another object, NGC~6337, which we describe in \Secref{Comparison}. 

Another concern is that the slit that measured the curved jet did not overlap its tip. This may work to our advantage, because the tip likely turns away in space and its velocity would suffer from an additional projection effect, which would not be easily quantified. As for determining the jet length, if the jet curved because of precession then we have indeed calculated a smaller jet length and underestimated its age. Similarly, if the angle of the jet is larger than 12 degrees, as assumed, then the deprojected velocity should be smaller and the jet older. It is possible that the time lag between CE ejection and jet be not so extreme, something that would help the interpretation of a post-CE jet as resulting from fall-back of material (\Secref{NGC6778Model}).  

As is the case for the other analysed PN, a final concern is that the structures were decelerated by ploughing up mass on their way. If the structures have been decelerated both their sizes and current velocities would be smaller than they should be. We ventured to guess that the jets would have been launched in a more evacuated environment since, as has been discovered by CE simulations (e.g., \citealt{Sandquist1998} and \citealt{Passy2012}), the CE ejection is equatorial. As a result, it is likely only a small amount of mass would be swept up. In addition, if the jet material had been significantly decelerated, then the original launch velocities would be higher, something that would be hard to reconcile with typical accretors encountered in PN. We argue here, as we have done in \Secref{TheNecklace} that any deceleration and mass loading should be within a factor of two and that this is supported by circumstantial evidence of the similarity of these jet speeds with those found in the other known post-CE jet PN, FLEMING~1 (\Secref{Comparison}).

\subsection{Conclusions from the data}
\seclabel{DataConclusions}

For the jets in the PN Abell 63, Necklace and ETHOS~1 we can say that:

\begin{enumerate}
  \item The jets predate the main nebula by $\sim\unit{5000}{\uyear}$.
  \item The jets have velocities of $\sim \unit{100}{\kms}$.
  \item For at least one central star (that of the Necklace PN), between $\unit{0.03 \mathand 0.45}{\Msun}$ were accreted onto the companion \citep{miszalski_carbon_2013}.
  \item For at least one of our three jet pairs, the mass is $\sim \unit{10^{-3}}{\Msun}$.
\end{enumerate}

As explained in \Secref{TheNecklace}, \Secref{Abell63} and \Secref{ETHOS1}, the uncertainties are large. However, we argue that the sequencing of jet and PN ejection is correct. We also argued that the jet speeds and ejected masses at the time of launch would not have been much higher than they are today. An additional cautionary note on the jet masses is that some gas may have recombined and was therefore not accounted for in our measurements. 

The mass loss rate of the jets is calculated using the jet mass estimate for the Necklace PN and assuming that the jet was launched for the entire dynamical age minus the dynamical age of the nebula, equivalent to assuming that the jet was launched continually up to when the ejection of the CE took place. This results in a jet lifetime between $\sim$4000 and $\sim$8000 years for the Necklace nebula, $\sim6000$~yr for A~63 and $\sim 5000$~yr for ETHOS~1. If the jet ejection timescales are lower, the jet mass-loss rates would be larger. We will further comment on this possibility in \Secref{AccretionAndEjectionMechanisms}.

The PN NGC~6778 is different in that:
\begin{enumerate}
  \item The jets formed after the main nebula by $\unit{\sim 3000}{\uyear}$.
  \item There are two pairs of jets.
  \item The velocities of both pairs of jets are higher $\unit{\sim 300-500}{\kms}$.
  \item The velocities of the two jet pairs are different.
\end{enumerate}
From these characteristics, we can already deduce that the jets from NGC~6778 are a post-CE event. The jet launch points are either closer to the central accretor(s) or the accretor(s) are more massive than for the pre-CE jet objects.  The jet mass loss rate is derived from the jet mass and a maximum jet lifetime of 1700 years.

It has to be emphasised that, although the formal errors could bring the age estimates of jets and ring to be much closer to one another, it is unlikely that the relative age estimates are completely unreliable. Looking at the data for the four objects in Table~1 and for the two additional objects which we discuss briefly in \Secref{Comparison}, we see a pattern, not only of relative ages, but also of jet speeds (slow for the pre-CE jets, faster for the post-CE jets). Although only improved measurements will refine this statement, it does appear that there are two distinct classes of CE jets. 

Below we consider three physical mechanisms for the accretion and ejection of mass: ejection of mass by radiative pressure (\secref{RadiativePressure}) and jet formation via an accretion disk formed at the time of Roche lobe (RL) overflow (\secref{RocheLobeOverflow}) or before RL overflow (\secref{BondiHoyle}). In order to be consistent with the calculations in each model, we have adopted the following parameter ranges:
\begin{itemize} 
  \item Mass of the jets: $M_{\text{jet}} \sim \unit{10^{-3}}{\Msun}$, this is based on the jet masses of Necklace PN and NGC~6778.
  \item Velocity of the jets: $v \approx \unit{100}{\kms}$ or $\unit{400}{\kms}$.
  \item Maximum duration of jet launching: $4000-8000$~yr for pre-CE jets, $\tau \sim \unit{1700}{\uyear}$ for post-CE ones.
  \item The mass-loss rate of the jets is $1-3 \times 10^{-7}$ or $8.8 \times 10^{-7} \Msun\usk\reciprocal\uyear$ for the pre and post-CE jets, respectively.
  \item Mass of the companion: $M_{\text{sec}} \sim \unit{0.3}{\Msun}$.
  \item Mass of the primary's core: $M_{\text{core}} \sim \unit{0.55}{\Msun}$.
  \item Mass of the primary's envelope: $M_{\text{env}} \sim \unit{0.45}{\Msun}$. 
\end{itemize}

The mass assumptions are appropriate for a $\unit{1.2}{\Msun}$ main sequence star, which is the median mass of the PN population \citep{Moe2006}. Such stars leave behind a $\sim0.55~\Msun$ core \citep{Weidemann2000,DeMarco2011}. At the time of interaction, the star is a giant and has a mass smaller than its main sequence mass. We therefore account for an envelope mass of 0.45~$\Msun$, so that our giant's total mass is 1~$\Msun$. The most represented stellar companion around white dwarfs has a spectral type M3.5V \citep{Farihi2005} which translates in a mass of $\sim0.3~\Msun$ \citep{DeMarco2013}.

\section{Accretion and Ejection Mechanisms for Jets from Common Envelope Systems}
\seclabel{AccretionAndEjectionMechanisms}

In our jet launching model, we assume that the accretion rate through the disk is $\sim$10 times the jet mass-loss rate derived in \Secref{DataConclusions} \citep{sheikhnezami_bipolar_2012}. Before we use these values of the accretion rate to derive the magnitude of the magnetic field (\Secref{magfields}), we consider the likely accretion rates in a series of probable accretion scenarios. Although many assumptions are made to derive values of accretion rates, such estimates provide one additional consistency check, which helps to gauge the reliability of the overall jet launch scenarios.

\subsection{Radiative Pressure}
\seclabel{RadiativePressure}

To calculate the radiative pressure exerted we use 
the brightest possible post-AGB star with $L \approx \unit{10^{4}}{L_{\odot}}$. Hence the radiative force, $\dot{p} = L/c$, is $\unit{10^{27}}{\dyne}$. This is the largest possible force, exerted if the entire radiation field of the star were intercepted by jet matter and converted to kinetic energy with maximum efficiency.

The smallest jet force in our sample is obtained for the pre-CE jets by using $M_{\rm jet} = \unit{10^{-3}}{\Msun}$ and velocity $v = \unit{100}{\kms}$ (\secref{DataConclusions}) with an accretion timescale of 8000 years. This results in $\dot{p}_{jet} \geq \unit{8 \times 10^{25}}{\dyne}$. A larger momentum limit can be obtained by using the fast jets of NGC~6778 (Table~1) with a mean jet velocity of $\unit{460}{\kms}$: $\dot{p}_{jet} \geq \unit{2 \times 10^{27}}{\dyne}$.

Comparing the lower limit range $8 \times 10^{25} - 2 \times 10^{27}$~dyne to the upper limit of $10^{27}$~dyne, it is easy to convince ourselves that radiation is unlikely to be responsible for the acceleration of these jets, not to mention that even if it were, there would be no explanation for the collimated nature of the outflows. We next turn our attention to accretion as a means to launch the jets.

\subsection{The accretion rate at the time of Roche lobe overflow}
\seclabel{RocheLobeOverflow}

The most logical moment to form an accretion disk in the life of a binary about to enter a CE interaction is at the time of RL overflow. To determine the mass accretion rate through the inner Lagrangian point we adopt, as a typical configuration, a $1~\Msun$ giant with a $300~\Rsun$ radius, entering RL contact with a $0.3~\Msun$ companion at 3~AU. Such system may be close to reaching synchronisation at the time of RL overflow.

A formalism for the accretion rate through the inner Lagrangian point is given by \citet{Ritter1988}\footnote{\citet{Davis2013} also provide a similar derivation; however, there are inconsistencies between these two papers. The formulae we use follow \citet{Ritter1988}, but correct a mistake in the algebra where the exponent ${-3/2}$ in their equation A8 should have been $-3$. \citet{Davis2013} have the correct exponent in their equation 9, but in the same equation they have multiplied by $q$, which is absent in the derivation of \citet{Ritter1988}. It should be noted, however, that the effect of the differences between the expressions used by \citet{Davis2013} and by \citet{Ritter1988}  was minimal.}. They define the accretion rate as:

\begin{equation}
  \dot{M} = \frac{2\pi}{\sqrt{e}} F\left(q\right) \frac{R_{L1}^{3}}{GM_{1}} \left(\frac{\mathcal{R} T_{\text{eff,1}}}{\mu_{\text{ph,1}}}\right)^{\frac{3}{2}} \rho_{\text{ph,1}},
  \eqlabel{RocheDavis}
\end{equation}
 
\noindent where $R_{L1}$ is the RL radius of the donor, in our case the primary giant, $\mathcal{R}$ is the ideal gas
constant, $G$ is the gravitational constant, $M_{1}$ is the mass of
the primary, $T_{\text{eff,1}}$ the effective temperature of the
primary, $\mu_{\text{ph,1}}$ is the mean molecular weight of the primary's atmosphere and
$\rho_{\text{ph,1}}$ is the density at the photosphere. $F\left(q\right)$ is
defined as:

\begin{equation}
  F\left(q\right) = \left(g\left(q\right)\left[g\left(q\right) - q - 1\right]\right)^{-1/2} \left(\frac{R_{L}}{a}\right)^{-3}
  \eqlabel{roche_F_func}
\end{equation}

\noindent with $g(q) = q/{x^{3}} + 1/(1 - x)^{3}$ and where $x$ is the distance of $L_{1}$  to the secondary in units of $a$. The value for $x$ can be calculated numerically based upon the orbital parameters of the system \citep[][figure 4]{Sepinsky2007}. The value of $x$ is 0.40 or $1.35 \times R_{L2}/a$. We also used $T_{\text{eff,1}} \sim\unit{3000}{\kelvin}$, $\mu_{\text{ph,1}} \sim 0.8$ (appropriate for a neutral cosmic mix), and $\rho_{\text{ph,1}} \sim\unit{10^{-9}}{\gram\usk\centi\metre\rpcubed}$ (appropriate for our AGB star's atmosphere). Finally, from \citet{eggleton_approximations_1983} the equation for the unit-less Roche radius of the primary is:
\begin{equation}
  r_{L1} = \frac{R_{L1}}{a} = \frac{0.49 q^{-\frac{2}{3}}}{0.6 q^{-\frac{2}{3}} + \ln\left(1 + q^{-\frac{1}{3}} \right)} 
  \label{eq:RocheRadius}
\end{equation}
where $q = M_{\rm sec}/M_{\rm prim}$, $R_{L1}$ is the RL radius for the primary, donor star and $a$ is the separation between the two objects. In this way we found $\dot{M} =\unit{4-8 \times10^{-4}}{\Msun\usk\reciprocal\uyear}$. We conclude that the RL overflow generates an accretion rate that may generate jets with a higher mass-loss rate and overall larger masses. This could either indicate that the pre-CE jets are launched before RL contact, or that the time over which they were launched is smaller, of the order of several decades, compared to the timescales we have adopted.

Finally, we remark that for an accretion rate of $8 \times 10^{-4}$~$\Msun$~yr$^{-1}$ and an accreted mass of $0.4~\Msun$ (see \Secref{TheNecklace}), the accretion timescale would be 500 years. This timescale is lower by a factor of a few than the maximum ejection timescales listed in Table~1. As scenarios and data are refined, all these timescales need to be reconciled.

\subsection{Wind Accretion}
\seclabel{BondiHoyle}

The issue of how long the system remains in RL contact before the onset of a CE is  important, because the more of the envelope is transferred to the companion the easier it will be for the companion to unbind the remaining envelope during the CE phase (both because the envelope is lighter and because the companion is more massive). In Section 3.2 we have concluded that the phase may not be particularly short because we have assumed that the mass accretion that gives rise to the observed jets takes place due to RL overflow. Here we consider the possibility that the accretion takes place before RL overflow via accretion from the wind of the primary. If this took place it may release the constraint of needing much accretion to take place at the time of RL overflow.

The Bondi-Hoyle mass accretion approximation \citep[BH;][]{bondi_mechanism_1944} cannot be used to investigate accretion at separations of a few AU, or just outside the RL overflow separation. 
The BH capture radius, $b=2GM/ (v_{\rm Kep}^2 + v_{\rm wind}^2)^{1/2}$, where the Keplerian ($v_{\rm Kep}$) and wind ($v_{\rm wind}$) velocities are similar, is of the order of the orbital separation. In such case the BH approximation cannot be valid since the medium through which the accretor/companion is moving is all but homogeneous.

Unfortunately, there are no analytical accretion models for the region just beyond RL contact. However, a few simulations have explored this region, either through the simulation of both the primary and secondary \citep{mohamed_shaping_2007,mohamed_wind_2011,kim_wide_2012}, or considering only one star \citep{huarte-espinosa_formation_2012}.  \citet{mohamed_shaping_2007,mohamed_wind_2011} propose an intermediate accretion mechanism that they call ``wind RL overflow'', where instead of the envelope of the primary filling the RL, its wind is channelled through the inner Lagrange point, allowing for an accretion rate that can be as high as half the mass-loss rate of the primary ($10^{-6} - 10^{-4}~\Msun$~yr$^{-1}$, for upper AGB stars) and much higher than the typical BH efficiency of a few percent. 
However, wind RL overflow may not always be applicable, as it requires that the velocity of the AGB wind at the RL radius be less than the escape velocity from the same location which will be sensitive to the details of the wind acceleration model. 

\citet{huarte-espinosa_formation_2012} simulated disk formation around a mass in a box with uniform fluid. Their disk mass for simulation setups that represented orbital separations between 10 and 20~AU ranged between $7 \times 10^{-6}$ and $6.5 \times 10^{-7}~\Msun$. Such disk masses would be  on the low side to explain our jets masses. However, better measurements and models that cover a wider parameter space, such as a smaller orbital separation may find some agreement.

\subsection{Accretion during the CE dynamical phase}
\seclabel{CEAccretion}

\citet{Ricker2008} and \citet{Ricker2012} determined the accretion rate onto the companion during the early phase of the fast dynamical inspiral. Although they conclude that the Bondi-Hoyle prescription would lead to an overestimate of the accretion rate by a factor of $\sim$100, their estimated average mass accretion rates are of the order of $10^{-2}$~M$_\odot$~yr$^{-1}$, which is a large value in the present context. In fact such rates would lead to jets with much larger accretion rates than we have measured and the only way to reconcile the numbers would be if the accretion timescales were lower than we have considered leading to larger jet mass-loss rates.

There are several issues with the estimate of \citet{Ricker2008} and \citet{Ricker2012} that prevent us from simply using their figure. First of all, from the scenario point of view a jet that developed because of accretion during the dynamical infall phase of the CE would have approximately the same dynamical age as the main nebula, something that is not observed in any of our systems, unless by some fluke of nature the uncertainties conspired in masking this coevality in all systems (\secref{data}). 

Secondly, these estimates are the result of a series of approximations, because the hydrodynamic simulations are not adequate to reproduce and hence measure the physics involved. The simulations do not model the surface of the accretor, but use instead a series of nested control surfaces. It is not clear that the mass that enters the control surfaces around the accreting companion actually accretes. This will depend on the angular momentum involved. The control surfaces also give discrepant rates, as is understandable, with the larger ones leading to higher accretion rates. Finally, such high accretion rates would be super-Eddington, which is also not included in the simulation.

This said, accretion during the CE phase would be a natural way to explain two jets since both the core of the giant and the companion may accrete material from the CE. It remains a priority of CE hydrodynamic simulations to establish when and how much accretion takes place.

\section{The magnetic field}
\seclabel{magfields}

Once an accretion disk forms, we require a mechanism to cause the material in the disk to lose angular momentum and launch the jets. The mechanism for the angular momentum loss that allows material to accrete onto the central object is a matter of debate \citep[see, e.g.,][]{stone_magnetohydrodynamics_1997,vishniac_angular_1993,gammie_theoretical_2005}. Here we assume that the angular momentum transport is provided by the magnetic field that is also responsible for launching the jets.  This in turn allows us to use the magnitude of the accretion rate to estimate the magnetic field strength.

\citet{wardle_magnetic_2007} derived an estimate of field strength required in order to accrete given the radius of the disk and the accretion rate (cf. section 2.1 of \citealt{Blackman2001}). The derivation considers the azimuthal component of the momentum equation for the system, and assumes that the azimuthal component of the disk's velocity can be approximated by its Keplerian velocity, given the disk is thin. Under such conditions, we have a minimum magnetic field strength (in Gauss) of:
\begin{equation}
  B \approx 0.2 \ \dot{M}_{-7}^{1/2} \ {r_{\text{AU}}^{-5/4}}\left(\frac{M}{\rm M_\odot}\right)^{1/4},
  \eqlabel{MagneticAccrete}
\end{equation}
where $M$ is the mass of the accretor, where $\dot{M}_{-7}$ is the mass accretion rate in units of $\unit{10^{-7}}{\Msun\usk \reciprocal\uyear}$ and $r_{\rm AU}$ is the disk radius, in AU, at which the field has that strength. This formalism applies locally, meaning that for a disk with an inner and an outer radii, one would derive a range of values of the magnetic field strengths. Below we apply this approximation to derive the magnitude of the magnetic field in our systems. We emphasise here that while the formalism above is a reasonable predictor of the needed magnetic field strengths, the accretion rates and disk sizes are not well constrained. However, as observationally-derived quantities become better known (likely from a larger sample, rather than better measurements) PN observations should put more stringent constraints on the CE interaction. 

\subsection{The magnetic fields in systems where the jets predate the nebula}
\seclabel{prePNfields}

\begin{table*}
  \caption{Minimum field strengths required to launch jets in the systems where jets predate the main nebula\tablabel{fields}}
  \begin{tabular}{lllllll}
     & Approximate  & Accretion & Field strength for & Field strength for & Field strength for & Field strength for \\
     & separation & rate & $r_{\text disk} = \unit{0.1}{\astronomicalunit}$ & $r_{\text disk} = \unit{0.5}{\astronomicalunit}$ & $r_{\text disk} = \unit{1}{\astronomicalunit}$ & $r_{\text disk} = \unit{2}{\astronomicalunit}$\\
    Mechanism & $\left(\astronomicalunit\right)$ & $\left(\Msun\usk\reciprocal\uyear\right)$ & $\left(\gauss\right)$ & $\left(\gauss\right)$ & $\left(\gauss\right)$ & $\left(\gauss\right)$\\
    \hline
    Wind accretion   & $\sim 3-5$    & $10^{-6} - 10^{-5}$ & $8 - 26$ &$1.1 - 3.5$ & $0.5 - 1.5$ & $0.2 - 0.5$ \\
    RL overflow        & $\sim 2-3$  &  $10^{-4} - 10^{-3}$ & $83 - 260$ &$11 - 35$ & $5 - 15$ & $2 - 6$ \\
    \hline
  \end{tabular}
\end{table*}
 
\citet{huarte-espinosa_formation_2012}, modelling wind accretion onto a companion orbiting at $\unit{10, 15 \mathand 20}{\astronomicalunit}$ from the primary, obtained accretion disk sizes of $\unit{\sim 1}{\astronomicalunit}$. 
A similar estimate for the disk radius is obtained in the case of RL overflow. Using the tidal equations of Zahn (1989) and the radius evolution of stars in the mass range $1-4~\Msun$, the maximum separation for a tidal capture is $5-8$~AU, but for the more common lower mass stars it is closer to $2-3$~AU \citep[e.g.,][]{Villaver2009,Mustill2012}, which is also the distance at which our typical $\unit{300}{\Rsun}$ giant will fill its RL. For a separation of 2.5~AU and a mass ratio of $M_2/M_1 = 0.3$, as adopted previously, we therefore expect the accretion disk radius to be smaller than the accretor's RL radius, or smaller than about 0.7~AU. \Tabref{fields} shows the required field strength using \Eqref{MagneticAccrete} with a range of accretion rates appropriate for wind and RL overflow accretion which also encompass the values deduced from the jet mass-loss rates (Table~1) and for accretion disk radii of 0.5, 1 and 2~AU which encompass likely values of such disks. We emphasise that these disk radii are to be interpreted as distances from the accretor where disk material would be losing angular momentum at a rate dictated by the local magnetic field and at which point a certain fraction of that material, assumed to be 10\%, would be launched vertically into a jet. Therefore the disk may extend to smaller and larger radii than the radius considered, but the magnetic field strength derived is for that location in the disk.

We finally note that the escape velocity from the gravitational field of accretors with masses between 0.3 and 1~$\Msun$, from a point located between 0.5 and 2~AU from the centre of the accreting secondary, are in the range 16-60~km~s$^{-1}$. These are lower than the jet speeds of $\sim$100~km~s$^{-1}$ measured for systems where the jets predate the CE ejection. To obtain such larger jet velocities, we would need disk radii of 0.05-0.2 AU or 10-45~$\Rsun$. So if we adopted a purely empirical approach, where we took a disk radius based on the jet speeds and an accretion rate of $2\times10^{-6}$~$\Msun$~yr$^{-1}$ (or ten times the jet mass-loss rate from Table~1), we would derive magnetic field strengths of 5 - 30 G (see Table~2).

\subsection{The magnetic field in NGC~6778: an indirect measurement of a post-CE magnetic field}

For the post CE PN NGC~6778 we know with reasonable certainty that the jets were launched after the CE dynamical infall phase. The two pairs of jets appear to be kinematically distinct so we also infer that they are not an optical illusion, part of the same kinematic structure under specific illumination conditions (as is the case for M2-9; \citealt{livio_twin_2001}). It is however difficult to construct a physical scenario for the launching of these jets because the post CE orbital separation leaves but a small space within which to form a sufficiently massive accretion disk (but see \Secref{NGC6778Model}).

However, on the assumption that these post-CE jets are indeed launched by a disk, we use their mass loss rate to infer a lower limit on the strength of the magnetic field necessary using \Eqref{MagneticAccrete}. Using the mass loss rate in the jets of $\unit{8.8 \times 10^{-7}}{\Msun\usk\reciprocal\uyear}$ (Table~1) and assuming, as we have done throughout this paper, that the accretion rate must be 10 times higher, and for a disk radius of $1$ and $10~\Rsun$, we obtain a magnetic field of $\unit{80 - 1400}{\gauss}$ (see \Tabref{ngcfields}). The orbital period of the binary today (Table~1), implies an orbital separation of approximately 1~$\Rsun$ for any plausible range of stellar masses. We include a larger disk radius in \Tabref{ngcfields} to encompass the possibility that at the time of the jet launching either the orbital separation was larger or that the disk was circumbinary (see \Secref{NGC6778Model}). We note that for an accretor in the mass range $0.3-1.0~\Msun$ and a launch point between 1 and 10~$\Rsun$, the escape velocity ranges between 110 and 620~km~s$^{-1}$, a range encompassing the deprojected velocities measured for the jets of NGC~6778 (Table~1).

We leave speculation of the actual scenario that gave rise to the post-CE twin jets to \Secref{NGC6778Model}.

\subsection{The origin of the magnetic field}

For the three systems where the jets predate the nebula, Gauss-sized magnetic fields are implied. These fields should thread the disk approximately vertically for the launching to happen according to commonly-adopted jet launching models \citep{Blandford1982}. The magnetic field could originate in the envelope of the giant as it gets spun up by the tidally-infalling companion. The field would have to be dragged as the envelope material moves towards the companion. It is less likely that the magnetic field would originate in the companion itself,  because of the relative old age of post-AGB binaries. One may also speculate that the field may be somehow self generated in the disk itself (for a discussion on different field configurations see \citealt{Pudritz2007}). Magnetic fields strengths on the surface of Miras have been measured and are consistent with a few Gauss. For example \citet{Amiri2012} measured a field strength of 3.5 G at 5.4~AU from the centre of the Mira star OH44.8-2.3.

For NGC~6778, whose launch model is so uncertain, it is paradoxically easier to hypothesise that the strong fields are created during the dynamical phase of the CE interaction, as detailed by \citet{regos_effect_1995} and \citet{Nordhaus2007}.  \citet{Nordhaus2007} modelled a CE dynamo in a $\unit{3}{\Msun}$ primary with secondaries in the mass range $0.02 - 0.05~\Msun$. They find that the toroidal field, $B_{\phi} \approx \unit{1-2 \times 10^{5}}{\gauss}$, while the poloidal component, responsible for the jet launching, $B_{\rm p} \approx \unit{200-300}{\gauss}$.

\Citet{regos_effect_1995}, provide an analytical formalism which we use here to derive the magnetic field components $B_{\phi}$ and $B_{p}$. We start with the equation for the poloidal component of the magnetic field \citep[equation 2.14]{regos_effect_1995}:
\begin{equation*}
 B_{\rm p} = 10 \gamma \left(\frac{3 M_{\text{env}}}{R_{\text{env}}^{3}}\right)^{1/2} \left(\frac{L R_{\text{env}}}{\eta M_{\text{T}}}\right)^{1/3},
\end{equation*}
where the efficiency of the dynamo regeneration term is $\gamma \sim 10^{-2}$, $\eta = 3R_{\text env}/l_{\rm c} \sim 30$, $R_{\text{env}}$ is the radius of the base of the envelope, $l_{\rm c}$ is the mixing length parameter, $L$ is the total energy generated in the envelope, and $M_{\text{T}} = M_{\text{env}} + M_{\text{R}} + M_{\text{W}}$, where $M_{\text{env}}$ is the mass of the envelope, $M_{\text{R}}$ is the mass of the secondary (they ``red" star), and $M_{\text{W}}$ is the mass of the core (the future white dwarf). The total luminosity is $L = L_{\text{stars}} + L_{\text{orb}}$, where:
\begin{equation*}
 L_{\text{stars}} = \frac{1.869 \times 10^{4} M_{\text{W}}^{7} +
7.205 \times 10^{6} M_{\text{W}}^{9}}{1 + 7.543 \times 10^{2}
M_{\text{W}}^{5} + 1.803 \times 10^{2} M_{\text{W}}^{7}},
\end{equation*}
\Citep[][equation 5.1]{regos_effect_1995} with $L_{\text{stars}}$ and $M_{\text{W}}$ in solar units. By combining their equations 4.6 and 4.9 to 4.11 and solving the resulting quadratic equation, we derive:
\begin{equation}
 L_{\text{orb}} = \frac{L_{\text{stars}}\left(1 + \sqrt{1 + 4a}\right)}{2a},
 \eqlabel{Lorbfunc}
\end{equation}
with:
\begin{equation*}
 a = \frac{\eta M_{\text{T}} L_{\text{stars}}}{\Delta\Omega}
\left(\frac{10}{M_{\text{env}} R_{\text{env}}
\Omega_{\text{orb}}}\right)^{2},
\end{equation*}
where $\Delta\Omega = \left|\Omega_{\text{orb}} - \Omega_{\text{env}}\right|$ is the difference (or shear) between the
angular velocity of the orbit ($\Omega_{\text{orb}} = \sqrt{{G(M_{\text{W}} + M_{\text{R}})}/{d^{3}}}$, where $d$ is the orbital separation), and that of 
the envelope. We read the value of the shear from figure 5 of \citet{regos_effect_1995}.
Finally, to determine $B_{\phi}$, we use their equation 2.4:
$ B_{\phi} = B_{p}/{\epsilon}$, with
\begin{equation*}
 \epsilon = \sqrt{\frac{0.01\left(\frac{R_{\text{env}} L}{\eta
M_{\text{T}}}\right)^{1/3}}{\Delta\Omega R_{\text{env}}}}.
\end{equation*}

Using a primary composed of an envelope with mass ranging between
$\unit{0.5 \mathand 2.5}{\Msun}$ and a core ranging between $\unit{0.5
\mathand 1}{\Msun}$, and a secondary with mass ranging between $\unit{0.5
\mathand 1.5}{\Msun}$, we found that $B_{\phi} \approx \unit{0.5-1\times 10^{4}}{\gauss}$ and $B_{p} \approx
\unit{100-500}{\gauss}$. Therefore \citet{regos_effect_1995} and
\citet{Nordhaus2007} agree on the magnitudes of the magnetic fields generated during a CE interaction. They both suggest that the field in post-CE primaries
would be mostly toroidal, but that their poloidal component is still relatively strong and similar to what we have determined using our jet observations, for the larger of the disk radii considered (\Tabref{ngcfields}). What we have not considered here is that the magnetic field would likely be transported out with the ejection of the CE, so that its strength at the location of the remnant binary would decrease in time.

\section{A Scenario for NGC~6778}
\seclabel{NGC6778Model}

\begin{table}
  \caption{Field strengths required to launch jets in NGC 6778\tablabel{ngcfields}}
  \begin{tabular}{p{1.7cm}p{2cm}p{2cm}}
    Accretion Rate $\left(\Msun\usk\reciprocal\uyear\right)$ & Field strength for $r_{\text disk} = \unit{1}{\Rsun}$ (G) & Field strength for $r_{\text disk} = \unit{10}{\Rsun} $ \\
    \hline
    $\sim 10^{-6}$ & $\unit{475}{\gauss}$ & $\unit{26}{\gauss}$ \\
    $\sim 10^{-5}$ & $\unit{1.6}{\kilo\gauss}$ & $\unit{92}{\gauss}$ \\
    $\sim 10^{-4}$ & $\unit{4}{\kilo\gauss}$ & $\unit{215}{\gauss}$ \\
    \hline
  \end{tabular}
\end{table}

Explaining the post-CE jet pairs observed in NGC~6778 is extremely difficult. At first sight, the two jets may indicate the formation of two accretion disks, possibly due to the infall of material that was not fully ejected by the CE interaction \citep[see, e.g.,][]{akashi_model_2008}. Such scenario may naturally explain the different jet velocities and even their morphologies. A simpler model whereby the jets are promoted by accretion of secondary star gas overfilling the RL and transferring to the primary via a disk may also be considered as was done by \citet{Soker1994b}. This is a plausible scenario since in a binary with only $\unit{1}{\Rsun}$ separation, the secondary star, with a radius of $\unit{0.5-1}{\Rsun}$ may indeed overflow its RL. 

In both scenarios the limited space between the two stars may limit excessively the mass of the disk that can form. In the RL-overflow scenario of \citet{Soker1994b} it would be hard to explain the two jets with their distinct kinematics. A third scenario already considered by \citet{Kashi2011} may be the formation of an accretion disk around both stars in the binary. However this would again not justify the two jet pairs. Additional scenarios may be constructed, for example one where the primary core spindown and strong magnetic fields expected after the CE ejection may form a jet in addition to one formed by a disk \citep{blackman_magnetohydrodynamic_2001}. However such scenario rely on complex physical mechanisms which may or may not be at play in these stars.

A further constraint on any scenario is the time between the CE ejection and the post-CE jet ejection ($\sim$3000~yr). Despite the difficulties mentioned above, we try here to determine whether a fallback disk would form within such a time frame. We here consider the infall of bound CE material and determine at what distance it would come to rest if we consider a simple ballistic trajectory and conservation of energy and angular momentum. By solving:

\begin{align}
  J_z &= m \ v_{\theta,h} \ h, \nonumber \\
  m \ v_{\theta,h} \ h &= m \ v_{\theta,disk} \ r_{disk}, \nonumber\\
  \frac{1}{2} v_{\theta, h}^2 - \frac{G M}{h} &=  \frac{1}{2} v_{\theta, disk}^2 - \frac{G M}{r_{disk}} 
\eqlabel{FallInRadius}
\end{align}
where $J_{z}$ is the angular momentum vector perpendicular to the orbital plane, $m$ is a mass element of in falling material, $v_{\theta,{h}}$, $v_{\theta,disk}$, $h$ and $r_{disk}$ are the orbital velocities and orbital radii of the material at altitude $h$ above the compact binary and at the altitude at which the disk comes to rest; $M$ is the mass of the central binary. 

In order to put some numbers into the solution, we refer to the CE simulations of \cite{passy_simulating_2012} for our estimates of the angular momentum of the infalling envelope gas and its initial distance from the central binary. The $z$ component of the total angular momentum of the system, $J_{z} \approx 2.5 \times10^{52}$~g~cm$^2$~s$^{-1}$,  was estimated by \citet{passy_simulating_2012} using a binary with $M_1 = 0.88~\Msun, M_2 = 0.6~\Msun\ {\rm and}\ a = 83 ~\Rsun$ (see their figure 8), where the angular momenta of the orbit and the envelope were considered. Of this, approximately $1/5$ belongs to bound matter \citep[see their figure 8]{passy_simulating_2012}. Bound material is distributed at  $h \sim 1 - 4.5 \times 10^3$~$\Rsun$ \citep[see their figure 19]{passy_simulating_2012}. If we divide the bound angular momentum by the mass of the bound envelope, using 95\% of an envelope of 0.49~$\Msun$, we get the value of the specific angular momentum of the infalling material: $v_{\theta,h} h \sim 5 \times 10^{18}$~cm~s$^{-1}$, which in turn gives $v_{\theta,h} \approx 1.5 - 7.2 \times 10^4$~cm~s$^{-1}$. Hence, using \Eqref{FallInRadius}, we calculate that the gas should come to rest at a distance from the binary centre of mass of approximately $\unit{0.06 - 20}{\Rsun}$. Hence, some of the fallback material will move closer to the centre of the binary than the orbital separation. Some of this material may have the correct angular momentum to form accretion disks around the two binary components, while some will be accreted directly onto the stellar surfaces or ejected from the system.

In order to determine the timescale of falling matter, we note that the gas follows half a Keplerian orbit with semi major axis $a = ( h + r_{disk}) / 2$, so by symmetry, the time taken is half the orbital period, or:  
\begin{equation*}
t = \pi \sqrt{a^3\over {GM}}. 
\end{equation*}
The result of this is the time taken for the mass element to fall to the equilibrium position, $t \sim 2-14$~yr. A more accurate ballistic calculation kindly carried out by J.-C. Passy using the results of his simulations \citep{passy_simulating_2012}, results in slightly longer timescales of $8 - 50$~yr, because this calculation accounts for the fact that the bound material is still carrying some outward velocity. 

Other physical mechanisms can be present that can slow down the infall: a fast wind from the central binary as well as radiation pressure. Using equation 6 from \citet{soker_backflow_2001}:
\begin{equation*}
  \frac{L/c}{\dot{M}_{\text{wind}}v_{\text{wind}}} = 10 \left(\frac{L}{\unit{5000}{\Lsun}}\right) \left(\frac{\dot{M}_{\text{wind}}}{\unit{10^{-8}}{\Msun\usk\reciprocal\uyear}}\right)^{-1} \left(\frac{v_{\text{wind}}}{\unit{1000}{\kms}}\right)^{-1},
\end{equation*}
where $L$ is the luminosity of the giant's core, $\dot{M}_{\rm wind}$ and $v_{\rm wind}$ are the mass-loss rate and velocity of the wind, respectively and $c$ is the speed of light. Here we have rescaled his values to those of an intermediate mass central stars (0.58~$\Msun$) transiting towards the white dwarf cooling track. The ram pressure of the wind is therefore a tenth of the radiation pressure and will not play a significant role in slowing down the infall.

To estimate the radiation pressure, we can again use \citet{soker_backflow_2001}. Using their equations 1, 2, 4 and 5, we can compute the ratio of the gravitational force to the radiative force, given by:
\begin{equation}
  \frac{f_{g}}{f_{r}} = \frac{\frac{G M m}{r^{2}}}{\frac{L}{c} \beta \left(1 - e^{-\tau}\right)}
  \eqlabel{force_ratio}
\end{equation}
where $m$ is a mass element at distance $r$ from the central binary with total mass $M$ and luminosity $L$, subtending a solid angle $\Omega$ such that $\beta = \Omega / 2 \pi$ and where $\tau$ is the optical depth of the mass element.
The optical depth of the infalling envelope is not easy to determine at present. However, even assuming that the material is optically thick, and that it subtends the entire solid angle, the gravitational force dominates by more than four orders of magnitude. Thus we deduce that radiation would not have much of a retarding effect on the infalling material. Lacking a way to retard the formation of a fall-back disk, we must conclude that such a short return timescale is at odds with the thousand-year-timescale indicated by the kinematics (unless the kinematic age of the jets were lower, or that of the disk higher).

\section{Comparison with CE PN Fleming~1 and NGC~6337}
\seclabel{Comparison}

There are two additional PN with jets known to harbour post-CE binaries: Fleming~1 and NGC~6337. We have not included them directly in our study because of the lack of nebular mass information. Below we review those characteristics which can be found in the literature and compare them to those of the 4 cases studied here.

Fleming~1 has jets that pre-date the nebula \citep{Lopez1993,Palmer1996}, as is the case for A~63, ETHOS~1 and the Necklace nebula. Fleming~1, with a 5000 year old main nebula and 16\,000 year old jets has the highest time interval between jet and CE formation. Its orbital period today is 1.19~days, similar to the Necklace nebula. Its flat jet caps are more similar to those of A~63. This nebula is thought to harbour a double degenerate star \citep{Boffin2012}. The deprojected fastest velocity of the knotty jets of Fleming~1 is $\sim$100~km~s$^{-1}$, assuming, as \citet{Boffin2012} have done, an inclination of 45~deg to the line of sight. This speed is in line with those of the other pre-CE jets.

NGC~6337, has post-CE jets as is the case for NGC~6778. The post-CE jets of NGC~6337 have many similarities with those of NGC~6778. Using the distance of Frew (2008) of 0.86$\pm$0.20~kpc (instead of the distance of 1.3~kpc of \citet{GarciaDiaz2009}), the ages of the nebula and jets are $\sim$8\,000 and $\sim$1000 years, respectively, a 7000 year delay between the CE and the jet ejection (cf. with almost 3000 years for NGC~6778). The jet velocity is $\sim$200~km~s$^{-1}$, smaller than the velocity of the jets of NGC~6778 (270 and 460~km~s$^{-1}$, for each of the two pairs), but larger than all the pre-CE jet speeds. The jets in NGC~6337 are bent as is one of the jet pairs in NGC~6778. The binary inside NGC~6337 has an orbital period of 0.17~days, similar to the very short period of the binary inside NGC~6778 (0.15~days). 

\citet{Hillwig2010} modelled the lightcurve of the central binary in NGC~6337 and, by assuming a central star mass of 0.6~$\Msun$, derived a companion mass of 0.2~$\Msun$ (quoting the hotter of the two models presented, but the differences are not large). This results in a situation where the companion, with a radius of 0.34~$\Msun$, is close to filling its RL (the inner Lagrangian point is only 0.56~$\Rsun$ away from the centre of the secondary). While we have considered a model where the disk is formed by re-accretion of nebular material, we must wonder whether the coincidence of both binaries with post-CE jets being so close to RL overflow may not be telling us that the jet is actually due to accretion of secondary material onto the primary as proposed by \citet{Soker1994b}. It is possible that in this case there differences in composition between the bulk of the nebula and the jet may be observed, since the jet may come from the unprocessed envelope of the main sequence secondary, rather than processed AGB envelope gas. Of course, this would be the case only if the envelope of the secondary were not highly contaminated by AGB envelope material accreted during the CE phase.

\section{Conclusion and Discussion}
\seclabel{Conclusion}

We have analysed the jets and nebulae of four post-CE PN, starting with their masses and kinematics. Three of the PN, the Necklace, Abell 63 and ETHOS~1, have jets that predate the main nebula by a few thousand years. They may have arisen when an accretion disk formed around the companion at the time of RL overflow, although that may lead to accretion rates higher than needed to explain the observed jets. Alternatively the pre-CE jets may have formed before RL contact, from wind accretion, in which case accretion rates could be lower. The latter hypothesis is also more in line with the relatively long timescales of jet formation before the CE infall phase. Further studies of systems like this could enable their use as constraints on the pre-CE phase which is at the moment ill constrained.

The fourth PN, NGC~6778, has jets that lagged the main nebula by about 3000 years. The two pairs of jets with different velocities, both higher than the jet velocities in the pre-CE jets, are difficult to explain by any scenario. Appealing to RL overflow of the companion after the CE ejection makes sense in view of the very small orbital separation of today's binary. However, the two pairs of jets are then difficult to explain. A scenario where fallback of envelope material forms one or two accretion disks around the binary or its components meets with difficulties both due to the need to delay the disk formation and the fact that the orbital separation is small enough that forming two disks would be difficult. Despite these difficulties it is clear that post-CE jets (of which there is at least another one in the PN NGC~6337) will be useful in constraining future simulations of the CE interaction.

Independently of the scenario that formed the accretion disks, we have derived the strength of the magnetic field that launches the jets using the assumption that it removes angular momentum at the rate needed for accretion of material and launches the jets according to the mechanism of \citet{Blandford1982}. If so, the magnetic field strengths are of a few to  ten Gauss, for pre-CE jets and hundreds to a few kilo-Gauss in the case of  post-CE jets. While it is unclear how to bring Gauss-strength fields to the proximity of the companion in a pre-CE binary, the strength of the post-CE fields is in line with the independent theoretical predictions of post-CE fields by \citet{regos_effect_1995} and \citet{Nordhaus2007}.

Finally we remark that the jet masses and kinematics can provide us with the indication of how much envelope has been ejected before the CE via the jets, and how much has accreted to the companion. Both these phenomena will facilitate the envelope ejection, something that could explain the lack of a full CE ejection witnessed in the simulations of \citep{passy_simulating_2012}.  Frew (2008) found that all CE PN have low ionised masses compared to the masses of other PN. This observation could be in line with the hypothesis of a decreased envelope mass. 

Accretion onto the companion is supported by the observation of carbon-rich material on the secondary star in the post-CE central binary of the Necklace nebula \citep{miszalski_carbon_2013}. Accretion onto the companion during the dynamical infall phase may contribute \citep{Ricker2008,Ricker2012}, although their mass accretion rates of $\unit{10^{-2}}{\Msun\usk\reciprocal\uyear}$ is likely overestimated and we would then expect an episodic jet or pair of jets launched at approximately the same time as the rest of the CE. An appealing feature of such scenario would be that the magnetic field at the time of launch would be naturally large, having been wound by the inspiral, and because the CE would not have departed yet, weakening the field in the proximity of the binary.  Undoubtedly, upcoming hydrodynamic simulations of the CE phase will be used to explore further this possibility.

\section*{Acknowledgments} We are thankful to Martin Guerrero and Romano Corradi for sharing their observations and deriving for us jet masses. We are also indebted with David Frew for providing us with a homogeneous set of distances to these PN, which may have improved the comparison between them. Jean-Claude Passy is thanked for calculating the ballistic trajectories of his common envelope simulations and confirming the analytical estimates of the fallback disks. Overall we are grateful for Noam Soker's extensive comments and criticisms. We are thankful to Jan Staff for sharing his theoretical knowledge of jets. Finally, we thank an anonymous referee for comments which allowed us to improve the paper. OD acknowledges Australian Research Council Discovery grant DP120103337 and Future Fellowship grant FT120100452; MJW acknowledges Discovery grant DP120101792.

\bibliographystyle{mn2e}

\bibliography{bibliography/AccretionDisks,bibliography/Bondi-Hoyle,bibliography/Jets,bibliography/MagneticFieldsDynamos,bibliography/Observations,bibliography/Other,bibliography/RocheLobes,bibliography/Orsola}
\end{document}